\documentclass[12pt]{iopart}

\usepackage{graphicx}
\usepackage{iopams}
\usepackage{color}
\usepackage{bm}

\newcommand{\ket}[1]{\left|#1\right>}
\newcommand{\bra}[1]{\left<#1\right|}

\newcommand{\nn}{\nonumber\\}

\newcommand{\bea}{\begin{eqnarray}}
\newcommand{\ea}{\end{eqnarray}}
\newcommand{\eea}{\end{eqnarray}}

\begin{document}

\title{Reverse quantum state engineering using electronic feedback loops}

\author{Gerold Kie{\ss}lich, Clive Emary, Gernot Schaller, and Tobias Brandes}

\address{Institut f\"ur Theoretische Physik, Technische
  Universit\"at Berlin, Hardenbergstra{\ss}e
  36,  D-10623 Berlin, Germany}

\ead{gerold.kiesslich@tu-berlin.de}

\begin{abstract}
We propose an all-electronic technique to manipulate and control interacting
quantum systems by unitary single-jump feedback conditioned on the outcome of a  capacitively coupled electrometer and in particular a 
single-electron transistor.
We provide a general scheme to stabilize pure states in the quantum system and employ an effective Hamiltonian method for the quantum master
equation to elaborate on the nature of stabilizable states and the
conditions under which state purification can be achieved. 
The state engineering within the quantum feedback scheme is shown to be linked with the solution of an inverse eigenvalue problem.
Two applications of the feedback scheme are presented in detail:
(i) stabilization of delocalized pure states in a single charge qubit
and (ii) entanglement stabilization in two coupled charge qubits.    
In the latter example we demonstrate the stabilization of a maximally entangled Bell state for certain detector positions and {\em local} feedback operations.
\end{abstract}


\submitto{\NJP}



\section{Introduction}

Quantum feedback control  is a promising scheme for the targeted manipulation of single quantum systems in which information gained from
a detector, which monitors a system, is used to direct appropriate control forces acting back on the quantum system \cite{WIS94,KOR99,KOR01,RUS02,KOR05,COM06,
JAC07,COM08,WIS10,RAL11}.

In solid-state systems some experimental realizations of quantum
feedback control schemes has been reported recently \cite{BLU10,GIL10,VIJ12,RIS12}. 
These examples have in common that the quantum system consists of a
quantum two-level system (qubit) and the feedback loops are realized 
all-electronically.
Different types of qubits are in use: 
Based on the proposal by Loss and di Vincenzo \cite{LOS98} {\it spin
qubits} consist of double quantum dots, where each dot is filled with
one electron \cite{FOL09,VAN11};
the two levels are represented by the singlet and one triplet of
the two-electron state.
A widely used class of qubits utilizes the charge degree-of-freedom of electrons.
In superconducting {\it charge qubits} the presence and absence of excess Cooper pairs on a
       superconducting island form the two-state system \cite{NAK99}.
A drawback of this design is its sensitivity to charge noise whereas the
{\it transmon qubit} provides an improved version  \cite{KOC07,SCH08b}.
Another charge qubit set-up in use is the normal-conducting double quantum dot,
where the excess electron can occupy either of the dots forming a
two-level system \cite{PET10}.
The latter type of quantum system will be the subject of our present studies. 

In all above cases the current qubit state is read out by a capacitively
coupled charge detector (electrometer), where the current flow through
the detector sensitively depends on the qubit state \cite{ASH09,ASH09a}.
The quantum point contact 
is the ``work horse'' among the charge detectors, 
since it provides a broad linear working range.
However, in this work we propose the two-state single-electron transistor (SET)
as charge detector for three reasons: (i) its high sensitivity, (ii) its low dimensionality 
(two charge states), and 
(iii) singular tunneling events to trigger feedback operations.
Other detector versions are the metallic SET \cite{WEI97} or the
radio-frequency SET \cite{SCH98a}, which is used in
superconducting devices. 

A manifold of objectives for quantum feedback control schemes is conceivable.  
The reduction of decoherence \cite{BLU10}, generation of
persistent quantum coherent oscillations \cite{VIJ12}, noise reduction
in quantum transport \cite{BRA10}, realization of an electronic Maxwell
demon \cite{SCH11,AVE11,ESP12}, target state preparation
\cite{RIS12}, entanglement stabilization \cite{STO04,WAN05,CAR07,CAR08,LIU10}, or stabilization of pure states \cite{HOF98,WAN01,JOR06} provide some of them. 
The latter attracted some recent theoretical work \cite{POE11,KIE11,EMA12}, where the feedback is assumed to be triggered not  by photon emissions but by 
electron jumps -- this is
the main subject of our present work.

We perform a systematic theoretical study of the electronic
implementation of a feedback scheme acting on a quantum
system using a SET as detector.
Thereby the detector will be considered as part of the system, which is a non-standard treatment -- this enables the systematic investigation of detector back-action and detector-induced 
dissipation.
We will demonstrate that the stabilizable pure states are eigenstates of the
effective Hamiltonian, which is defined by the quantum master equation
for the coupled system-detector dynamics.
It will be shown that the feedback stabilization procedure defines an inverse eigenvalue
problem, which enables a more systematic way to obtain convenient
feedback operations.

In the spirit of the above mentioned experiments we will demonstrate the stabilization of pure states in a
single charge qubit for arbitrary system-detector couplings. 
Moreover, inspired by a recent experimental demonstration of entanglement of
electrostatically coupled singlet-triplet qubits \cite{SHU12} we study
the entanglement stabilization in two coupled charge qubits. 

The paper is organized as follows: In section~\ref{sec:engineer} we describe, what is meant by ``quantum state engineering'' and why our 
feedback control scheme provides a reverse technique.
Section~\ref{sec:general} contains the general derivation of the feedback scheme with the introduction of the microscopic model (\ref{sec:hamilt}), 
the introduction of the effective Hamiltonian approach (\ref{sec:effective}), and the formulation of the inverse eigenvalue problem (\ref{sec:feedback}).
In section~\ref{sec:examples} we provide two examples of quantum systems for illustration purposes: a single charge qubit (\ref{sec:qubit}) 
and two interacting charge qubits 
(\ref{sec:entangle}).
The conclusions can be found in section~\ref{sec:conclude}.


\begin{figure}[b]
\begin{center}
\includegraphics[width=0.8\textwidth]{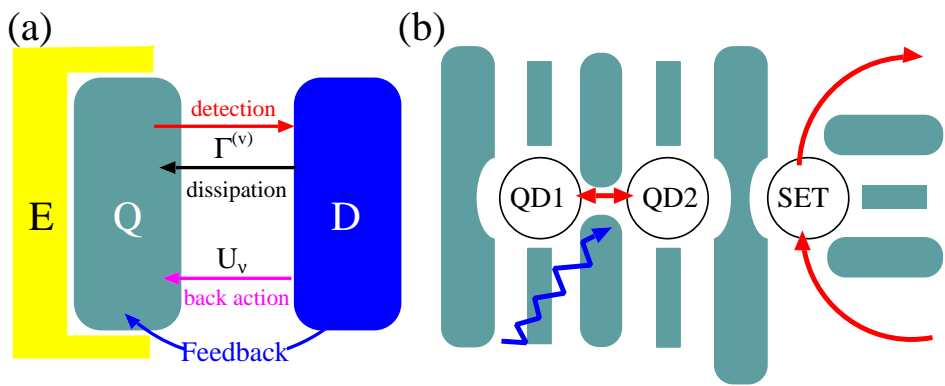}
\end{center}
\caption{(Color online) {\bf (a)} Schematic set-up of the electronic feedback device. (Q)uantum
system, (D)etector, (E)nvironment of $Q$. 
Note that we distinguish between the intrinsic back-action from the detector on $Q$ and the feedback signal, which undergoes a classical processing by the 
experimenter.
{\bf (b)} Device implementation with single charge qubit as quantum system 
and single-electron transistor as detector.
}
\label{fig:scheme}
\end{figure}


\section{\label{sec:engineer}Quantum state engineering}

The creation of quantum states which are temporally stable and robust against perturbations is a major requirement for the advancement of quantum-based technologies.
One main technical issue deals with the continuous state detection, which typically leads to the irreversible 
loss of quantum purity due to ongoing projective measurements.
But also the dissipative influence of the remaining  environment constitutes a source of decoherence, which one needs to avoid in the 
best case.  
However, at least the destructive (back-) action of the detector can be suppressed or even reversed, e.g., 
application of the properly processed detector signal on the 
quantum system as proposed theoretically by us for an all-electronic set-up \cite{KIE11}.
Beyond that, this feedback method
enables the stabilization of a set of pure quantum states depending on the specific feedback 
scheme and parameters (In contrast to back-action and dissipation that modify the dynamics of a quantum system in a unitary or 
non-unitary manner, respectively, feedback in our notion requires prior classical processing of the detector signal and may thus be modified by the 
experimenter at any time.).
So far this procedure has not really been systematic --  for 
application purposes in state engineering, however, a clearer instruction will be necessary.
In this work we provide such a sequence of steps that the experimenter or quantum engineer can follow in order to realize our feedback method.

The usual way to generate a quantum steady-state induces the solution of an eigenvalue problem for closed systems.
Alternatively, one solves, e.g., the steady-state version of the quantum master equation for dissipative systems.
Either the Hamiltonian operator or the Liouvillian super-operator is presumed to be known.
Our proposed feedback scheme starts with a given system-detector set-up and seeks the corresponding control action that drives the system to one of the allowed target states.
Similar reverse ans{\"a}tze can be found for dissipative quantum state engineering
\cite{RUB82,POY96,BEI00,VER09,DIE11,KOG12}, for projective measurements
\cite{VOL11}, and for feedback \cite{LLO01,TIC12}.

The following principle steps need to be performed:
\begin{itemize}
\item[(1)] What type of states can be stabilized ? $\Leftrightarrow$  
Compute the spectrum of the effective Hamiltonian [(\ref{eq:effHam}) in section~\ref{sec:effective}] of the system with detector and without control.
Note that the eigenstates of the effective Hamiltonian are not equal to those of the system Hamiltonian without detection. 

\item[(2)] Define an appropriate target state $\Leftrightarrow$  Choose one of the eigenstates. 
\item[(3)] Compute the feedback operation which yields the target $\Leftrightarrow$ Insert the eigenstate into an inverse eigenvalue problem and compute the corresponding
(feedback) super-operator [(\ref{eq:feedback}) in section~\ref{sec:feedback}]. 
\end{itemize}
In the next section we will provide the (mathematical and physical)
details of this procedure.


\section{\label{sec:general}Feedback scheme}

\subsection{\label{sec:hamilt}Hamiltonian}

The total Hamiltonian reads
\bea
\hat{H}=\hat{H}_Q+\hat{H}_E+\hat{H}_D+\hat{H}_C,
\eea
where $\hat{H}_Q$ describes a quantum system ($Q$) to be stabilized.
We will make a particular choice for our continuously operating detector:
we employ a single-electron transistor (SET) with its standard Hamiltonian
\bea
\hat{H}_D=\hat{H}_{\textrm{SET}}=\epsilon_d
\hat{d}^\dagger\hat{d}+\sum_{k,\alpha=S,D}\big[
\epsilon_{k\alpha}
\hat{c}_{k\alpha}^\dagger\hat{c}_{k\alpha}+(t_{k\alpha}\hat{c}_{k\alpha}^\dagger\hat{d}+\textrm{h.c.})\big],
\eea 
where $\hat{d}^\dagger / \hat{d}$ denotes the creation/annihilation operator of the SET state with level energy $\epsilon_d$.
The $k$-th mode of the electronic source/drain lead ($\alpha =S/D$) is created/annihilated with operator $\hat{c}_{k\alpha}^\dagger /\hat{c}_{k\alpha}$
and has the energy $\epsilon_{k\alpha}$.
The coupling between SET level and contact modes is given by the tunnel matrix element $t_{k\alpha}$.
The advantage of this choice for the theoretical treatment 
is that this type of detector  possesses only two degrees of freedom after tracing out the electronic leads.
The coupling between $Q$  and the SET is assumed to be
\bea
\hat{H}_C=\hat{d}^\dagger\hat{d}\otimes \hat{X},
\eea
 where $\hat{X}$ consists of  operators acting solely in $Q$ (see below).

The term \mbox{$\hat{H}_E=\hat{H}_B+\hat{H}_I$} describes an environment of $Q$, where $\hat{H}_B$ is the bath Hamiltonian (e.g., phonons) and   $\hat{H}_I$ provides its interaction 
with $Q$.
For the sake of generality we will not specify it further at this stage.
%


\subsection{\label{sec:eom}Equation of Motion}

The standard Born-Markov approximation \cite{BRE02,SCH09c} for weak coupling of the SET 
in the infinite bias limit and of the quantum system with a bath 
yields the Lindblad-type equation of motion for the density matrix of the system ($\hbar =$~1):
\bea
\frac{d}{dt}\hat\rho (t)  = -i[\hat{H}_{\textrm{sys}},\hat\rho (t)]-\frac{1}{2}\sum_{\alpha
  =E/S/D}\bigg[\big\{\hat{L}_\alpha^\dagger\hat{L}_\alpha,\hat\rho (t)\big\}
-2\hat{L}_\alpha\hat\rho (t)\hat{L}_\alpha^\dagger\bigg],
\label{eq:lindblad}
\eea
where the system part is given by 
\bea
\hat{H}_{\textrm{sys}}=\hat{H}_Q+\epsilon_d\hat{d}^\dagger\hat{d}
+\hat{H}_C,
\eea
so that \mbox{$\textrm{dim}(\hat{H}_{\textrm{sys}})=2\,\textrm{dim}(\hat{H}_Q)$}, where the pre-factor 2 stems from the two SET states.
In the standard derivation the Lindblad operators for the SET tunneling are obtained as $\hat{L}_D = \sqrt{\Gamma_D}\,\mathbf{1}_{\textrm{sys}}\otimes\hat{d}$ and
$\hat{L}_S = \sqrt{\Gamma_S}\,\mathbf{1}_{\textrm{sys}}\otimes\hat{d}^\dagger$
with the tunneling rates
\bea
\Gamma_\alpha (\omega )=2\pi\sum_k\vert t_{k\alpha}\vert^2\delta (\omega -\epsilon_{k\alpha}),\quad \alpha\in\{S,D\}.
\eea
In the wide-band approximation and infinite bias limit the rates are energy independent $\Gamma_\alpha =\Gamma_\alpha (\omega )$, so that 
the SET is entirely decoupled from $Q$ except of the interaction $\hat{H}_C$ entering the first term on the right hand side of (\ref{eq:lindblad}).
It merely acts on the quantum system, but not on the SET.

For the following we provide the $Q$-part of the interaction between SET and quantum system in its spectral decomposition
\bea
\hat{X}=\sum
  U_\nu\ket{\nu}\bra{\nu},
\eea
which serves as definition of the states $\ket{\nu}$ and the interaction strengths $U_\nu$.
A specific realization of such a type of interaction may be the Coulomb interaction of the SET electron with  the electrons confined in the quantum system, 
which we are going to discuss in section~\ref{sec:examples}.

For the specification of the Lindblad operators in (\ref{eq:lindblad}) we now utilize the states $\ket{\nu}$
and follow the phenomenological approach of Refs.~ \cite{GUR96c,GUR08}, where the tunneling rates are conditioned on whether
state $\nu$ is occupied: $\Gamma_\alpha^{(\nu )}$.
In the above example of Coulomb interacting electrons one can argue that the repulsion of electrons induces an energy shift of the SET level 
 $\varepsilon_d\to\varepsilon_d + U_\nu$. 
With increasing $U_\nu$, it follows that the SET electrons experience lower tunneling barriers and correspondingly higher tunneling rates $\Gamma_\alpha$.
The Lindblad operators then turn out to be:
\bea
\hat{L}_D &=&\hat{B}_D\otimes\hat{d},\quad
\hat{L}_S =\hat{B}_S\otimes\hat{d}^\dagger\\
\textrm{with}&\quad&\hat{B}_\alpha\equiv\sum_{\nu}\sqrt{\Gamma_\alpha^{(\nu)}}
\ket{\nu}\bra{\nu}.\nonumber
\label{eq:lindbladop}
\eea
This leads to a dissipative coupling of the SET and the quantum system.

The interaction of the quantum system with an environment is contained
in the Lindblad operators $\hat{L}_E$ in (\ref{eq:lindblad}),
which will be specified later, e.g., in (\ref{eq:lindblad_env}).


\subsection{\label{sec:effective}Effective Hamiltonian}

The Markovian master equation  of Lindblad form (\ref{eq:lindblad}) can
be written as 
\bea
\frac{d}{dt}\hat\rho (t) =\mathcal{L}\hat\rho (t)
=\big(\mathcal{L}_0+\mathcal{J}\big)\hat\rho (t)
\label{eq:master}
\eea
with \mbox{$\mathcal{J}\hat\rho (t)$ $\hat{=}$ $\sum_{\alpha=E,S,D}\hat{L}_\alpha\hat\rho (t)\hat{L}_\alpha^\dagger$} for environment-induced jumps in
the quantum system and
electron jumps through the SET (throughout the following we will use
calligraphic symbols to denote super-operators).
The free Liouvillian $\mathcal{L}_0$ describes the evolution of the system without electron transfer
between sub-systems and reservoirs and assumes the form (see also \cite{POE11})
\bea
\mathcal{L}_0\hat\rho (t)=-i\{\hat{H}_{\textrm{eff}}\hat\rho (t)-\hat\rho (t)\hat{H}_{\textrm{eff}}^\dagger\},
\eea
where
\bea
\hat{H}_{\textrm{eff}}=\hat{H}_{\textrm{sys}}-\frac{i}{2}\sum_{\alpha=E,S,D}\hat{L}_{\alpha}^\dagger\hat{L}_\alpha
\label{eq:effHam}
\eea 
is an effective non-Hermitian Hamiltonian operator for the system.
Note that the effective Hamiltonian is invariant under unitary
transformations of the Lindblad operators $\hat{L}_\alpha$ and also
inhomogeneous shift transformations that leave the Lindblad form
invariant.
The effective Hamiltonian has right and left eigenstates 
$\hat{H}_{\textrm{eff}}\ket{\psi_k}=\epsilon_k\ket{\psi_k}$ and
$\tilde{\bra{\psi_k}}\hat{H}_{\textrm{eff}}=
\epsilon_k\tilde{\bra{\psi_k}}$, which, in general, are non-adjoint and
the eigenenergies are complex.
These states will be used to construct the eigenoperators of the free
Liouvillian, which are obtained by
\bea
\mathcal{L}_0\hat\rho_{jk}=-i(\epsilon_j-\epsilon_k^*)\hat\rho_{jk},
\eea
with $\hat\rho_{jk}=\ket{\psi_j}\tilde{\bra{\psi_k}}$.
The diagonal eigenoperators \mbox{$\hat{\rho}_k\equiv\hat\rho_{kk}$} obey 
\bea
\mathcal{L}_0\hat\rho_{k}=2\Im{(\epsilon_k)\hat\rho_{k}},
\eea
and represent pure states -- due to this property they will play a crucial role
in the following.

For our system we decompose the effective Hamiltonian
(\ref{eq:effHam}) 
with respect to the SET charge state $\ket{n}$ ($n\in\{0,1\}$) as
\bea
\hat{H}_{\textrm{eff}}&=&\sum_{n=0,1}\hat{H}_\textrm{eff}^{(n)}\otimes\ket{n}\bra{n},\quad
\textrm{where}\\
\hat{H}_\textrm{eff}^{(0)}&\equiv&\hat{H}_Q-\frac{i}{2}\bigg(\hat{L}_E^\dagger\hat{L}_E
+\hat{B}_S^\dagger
\hat{B}_S\bigg),\nn 
\hat{H}_\textrm{eff}^{(1)}&\equiv&\hat{H}_Q-\frac{i}{2}\bigg(\hat{L}_E^\dagger\hat{L}_E
+\hat{B}_D^\dagger
\hat{B}_D\bigg)+\sum_{\nu}U_{\nu}\ket{\nu}\bra{\nu}+\epsilon_d\mathbf{1}\nonumber
\label{eq:effHam2}
\eea
with $\hat{B}_\alpha^\dagger
\hat{B}_\alpha =\sum_{\nu}\Gamma_\alpha^{(\nu)}\ket{\nu}\bra{\nu}$.
It follows that the eigenstates and -energies can be separated with
respect to the
SET charge state: 
\bea
\hat
{H}_\textrm{eff}^{(n)}\ket{\psi_{nk}}=\epsilon_{nk}\ket{\psi_{nk}},\quad
\tilde{\bra{\psi_{nk}}}\hat{H}_\textrm{eff}^{(n)}=\epsilon_{nk}\tilde{\bra{\psi_{nk}}}.
\eea 

Our aim is the stabilization of a pure state in the quantum system so that we make use of theses eigenstates to build the operators 
\bea
\hat{\rho}_k^{(0)}=\ket{\psi_{0k}}\tilde{\bra{\psi_{0k}}},\quad\hat{\rho}_k^{(1)}=\ket{\psi_{1k}}
\tilde{\bra{\psi_{1k}}}
\eea
with $\textrm{Tr}\{[\hat{\rho}_l^{(0,1)}]^2\}=$~1.
Now, the state, which we seek to stabilize, we write in the
general form 
\bea
\hat{R}_{k,k'}=c_0\hat{\rho}_k^{(0)}\otimes\ket{0}\bra{0}+c_1\hat{\rho}_{k'}^{(1)}\otimes\ket{1}\bra{1},
\eea
where  due to normalization \mbox{$c_0+c_1=1$} holds.
Note that there are no coherences between different SET charge
states since these correspond to (forbidden) superpositions of states of different charge.
The corresponding state of the isolated quantum system is  
\bea
\hat{\rho}_{Q,k,k'}\equiv\textrm{Tr}_{\textrm{SET}}(\hat{R}_{k,k'})=
c_0\hat{\rho}_k^{(0)}+c_1\hat{\rho}_{k'}^{(1)}, 
\eea
which is a mixture in general.
To obtain a pure state for finite SET current (i.e., $c_i>$~0)
$\hat\rho_{Q,k}\equiv\hat{\rho}_k^{(0)}=\hat{\rho}_{k'}^{(1)}$ must be
fulfilled.
(Pure states also can be obtained for $c_0=$~1 or
  $c_1=$~1, but those do not correspond to a finite SET current. The
  feedback loop could not be closed in  those cases.)
With the help of (\ref{eq:effHam2}) it readily follows that
$\Gamma_S^{(\nu )}=\Gamma_D^{(\nu )}\equiv\Gamma^{(\nu )}/2$, $\epsilon_{0k}=\epsilon_{1k}\equiv\epsilon_k$, and $c_0=c_1=1/2$.
Furthermore, we need to demand that 
\bea
U_{\nu}\ll\textrm{max}\bigg[\bra{\nu}\hat{H}_Q\ket{\nu'},\Gamma_\alpha^{(\nu )}\bigg],
\label{eq:maxcond}
\eea
so that this
term can be neglected in $\hat{H}_{\textrm{eff}}^{(1)}$, but the differences
between the $\Gamma^{(\nu )}$ are still resolved.
Our desired state then will be a direct product  of quantum system and SET state: 
\bea
\hat{R}_k=\hat\rho_{Q,k}\otimes\hat\rho_{\textrm{SET}}
\label{eq:product}
\eea
with the steady-state mixture \mbox{$\hat\rho_{\textrm{SET}}
=\textrm{Diag}(1/2,1/2)$} of a symmetric SET.
Hence, in order to stabilize pure states in the isolated quantum
system a 
dis-entanglement between detector and quantum system has to be forced.
This can be achieved by a vanishing direct back-action $U_{\nu}$ from the
SET towards the quantum system [see figure~\ref{fig:scheme}(a)]
and symmetric tunnel coupling in the SET.


\subsection{\label{sec:feedback}Feedback stabilization}

The clicks obtained from a measurement of electron jumps at the source or drain barrier of the SET detector will be used to trigger short time pulses on the quantum
system Hamiltonian.
In experiments immediately after an electron jump is detected an electric voltage pulse will be applied at a 
metallic gate in the electronic device which belongs to $Q$ [Fig.~\ref{fig:scheme}(b)].

As shown in \ref{sec:appendix1} for unit detection efficiency this leads to a modification of the Markovian quantum master equation (\ref{eq:master})
\bea
\frac{d}{dt}\hat\rho (t)
=\bigg[\mathcal{L}_0+\mathcal{J}_E+\sum_{\alpha=S,D}\mathcal{C}_\alpha\mathcal{J}_\alpha\bigg]
\hat\rho (t).
\label{eq:master2}
\eea
The SET jump operators $\mathcal{J}_\alpha$ are supplemented by the unitary operations 
$\mathcal{C}_\alpha\hat\rho = e^{\mathcal{K}_\alpha}\hat\rho$
with the superoperators $\mathcal{K}_{S/D}$ given by 
\bea
\mathcal{K}_\alpha\hat\rho
  (t)=-\frac{i}{\hbar}\big[\hat{h}_\alpha,\hat\rho (t)\big]\delta t,
\eea
where $\hat h_\alpha$ acts on the quantum system during the time interval $\delta t$.
Such a feedback scheme has been introduced by Wiseman and Milburn in a
quantum optical context \cite{WIS94,WIS10}.

In order to realize this scheme, single electron jumps in the SET need to be resolved.
However, in an experimental set-up the sequence of single jump events at the SET barriers 
which corresponds to the respective current
$I_{\alpha}(t)=e\sum_k\delta (t-t_k^{(\alpha)})$ 
may neither be
resolved nor measured independently in the SET circuit.
One possible solution may be the implementation of a  quantum-point contact (QPC)
weakly attached to the SET as proposed in Ref.~\cite{KIE11}.

Inserting the operators (\ref{eq:product}) into the steady-state
version of (\ref{eq:master2}) yields
\bea
\hat{\mathcal{G}}^{(k)}\equiv\bigg[2\Im{(\epsilon_{k})}\mathbf{1}
+\mathcal{A}_E+\mathcal{C}\mathcal{A}\bigg]\hat{\rho}_{Q,k}&=&0,
\label{eq:feedback}
\eea
where $\mathcal{A}_E$ is defined by
$\mathcal{J}_E\big[\hat{\rho}_{Q,k}\otimes\hat{\rho}_{\textrm{SET}}\big]=\mathcal{A}_E\hat{\rho}_{Q,k}\otimes\hat{\rho}_{\textrm{SET}}$,
and $\mathcal{A}$ is defined by  
$\mathcal{J}_S\big[\hat{\rho}_{Q,k}\otimes\ket{0}\bra{0}\big]=\mathcal{A}\hat{\rho}_{Q,k}\otimes\ket{1}\bra{1}$
or $\mathcal{J}_D\big[\hat{\rho}_{Q,k}\otimes\ket{1}\bra{1}\big]=\mathcal{A}\hat{\rho}_{Q,k}\otimes\ket{0}\bra{0}$.
The feedback must be symmetric from the source and drain SET current, 
such that $\mathcal{C}\equiv\mathcal{C}_S=\mathcal{C}_D$.
(\ref{eq:feedback}) provides the central result of this work and defines an inverse eigenvalue problem, where the eigenenergies and -states are known to
belong to the effective Hamiltonian $\hat{H}_{\textrm{eff}}$.
The feedback super-operator $\mathcal{C}=e^{\mathcal{K}}$, in particular
$\hat{h}$, will be sought.
Solving such a problem can be considered as reverse state engineering, as already discussed in section~\ref{sec:engineer}  --
one needs to determine the feedback operation to stabilize a given
quantum state, which is chosen to be compatible with the detection scheme.
To provide a measure of stabilization quality we will use the
Hilbert-Schmidt norm
\bea  
g_k\equiv\sum_{ij}\big\vert\hat{\mathcal{G}}^{(k)}_{ij}\big\vert^2\ge
0
\label{eq:schmidt-hilbert}
\eea
of the left hand side of
(\ref{eq:feedback}); where equality holds for perfect
stabilization of the $k-$th eigenstate of the effective
Hamiltonian.


\section{\label{sec:examples}Applications of the feedback scheme}

In this section we provide two electronic examples, which will illustrate the general feedback scheme with the aim to  stabilize pure states.
The first example deals with a single-electron quantum system: a charge qubit.
Secondly, the stabilization of entangled states will be studied in a system of two interacting charge qubits.


\begin{figure}[t]
\begin{center}
\includegraphics[width=.62\textwidth]{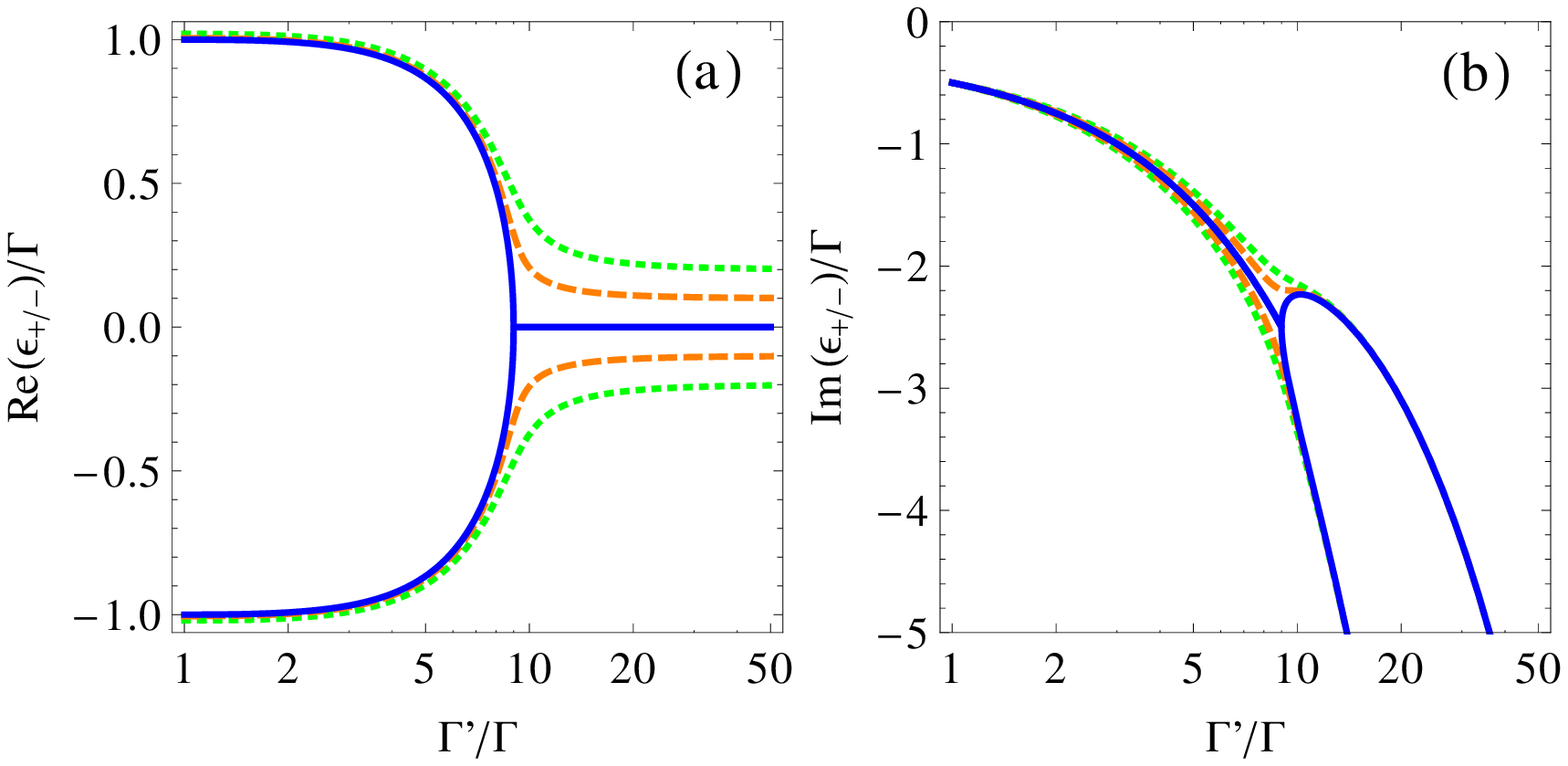}
\includegraphics[width=.32\textwidth]{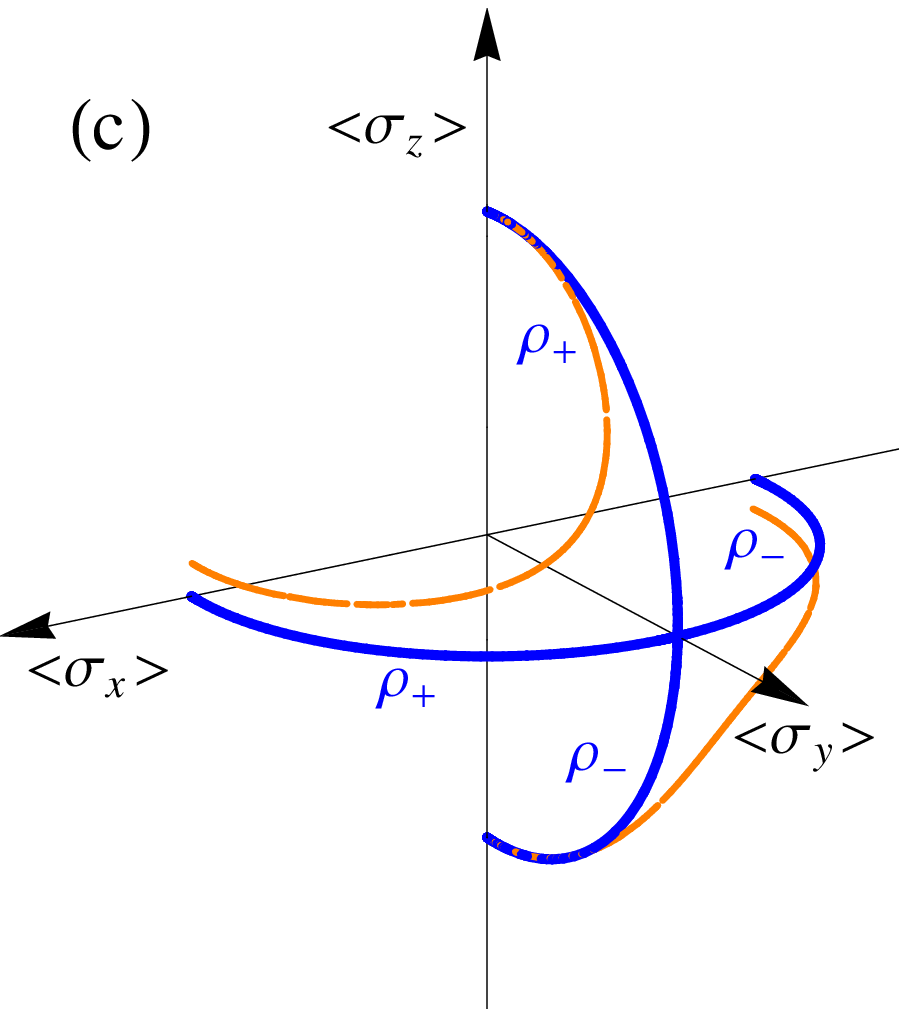}
\end{center}
\caption{(Color online) (a) Real part and (b) imaginary part of eigenenergies of the single qubit effective Hamiltonian
  (\ref{eq:effHam}); (blue, solid) $\varepsilon=$~0, (orange, dashed)
  $\varepsilon=$~0.2, (green, dotted)
  $\varepsilon=$~0.4. (c) Qubit Bloch vectors of eigenstates of effective Hamiltonian
  (\ref{eq:effHam}) for varying $\gamma_-$ and vanishing back action $U =$~0.
  $\vert\gamma_-\vert <4T_C$:
  delocalization. $\vert\gamma_-\vert >4T_C$:
  localization.  (blue, solid) $\varepsilon=$~0, (orange, dashed)
  \mbox{$\varepsilon=$~0.2}.}
\label{fig1}
\end{figure}


\subsection{\label{sec:qubit}Single-electron quantum system: Dissipative charge qubit}

\noindent
{\em Model.} -- The charge qubit Hamiltonian  reads
\bea
\hat{H}_Q=\frac{\varepsilon}{2}\hat{\sigma}_z+T_C\hat{\sigma}_x,
\eea
with the qubit bias $\varepsilon\equiv\varepsilon_t-\varepsilon_b$ and
the coupling between the dots $T_C$.
The Hamiltonian is given in the Bloch representation with Pauli matrices:
$\hat{\sigma}_z=\ket{t}\bra{t}-\ket{b}\bra{b}$ and 
$\hat{\sigma}_x=\ket{t}\bra{b}+\ket{b}\bra{t}$.
As a specific source of dissipation we consider 
background charge fluctuations where the qubit bias is fluctuating: 
$\varepsilon (t) = \varepsilon + \xi (t)/\sqrt{\tau}$ 
with
$\langle\xi (t)\rangle =$~0 and $\langle\xi (t)\xi (t')\rangle =\delta (t-t')$.
This yields the dissipative Lindblad operator \cite{KIE09}
\bea
\hat{L}_E=\frac{1}{\sqrt{\tau}}\hat{\sigma}_z\otimes\mathbf{1}_{\textrm{SET}}.
\label{eq:lindblad_env}
\eea
Similar dissipators also can be found for electron-phonon coupling in
the high temperature limit
\cite{BRA05}.

The SET couples capacitively with the qubit so that
$\hat{X}=\frac{U}{2}\hat{\sigma}_z$ ($U\equiv U_t-U_b$) and
the Lindblad operators in the master equation (\ref{eq:lindbladop}) become
\bea
\hat{L}_D &=& \bigg(\sqrt{\Gamma_D'}\ket{t}\bra{t} +\sqrt{\Gamma_D}\ket{b}\bra{b}\bigg)\otimes d,\nn
\hat{L}_S &=& \bigg(\sqrt{\Gamma_S'}\ket{t}\bra{t} +\sqrt{\Gamma_S}\ket{b}\bra{b}\bigg)\otimes d^\dagger .
\eea

\noindent
{\em Spectrum of the effective Hamiltonian.} -- Without dissipation the eigenenergies of the effective Hamiltonian (\ref{eq:effHam}) are:
\bea
\epsilon_{0\pm}=-i\frac{\gamma_S^+}{2}\pm\sqrt{e_S(\varepsilon)^2+T_C^2},\quad
\epsilon_{1\pm}=-i\frac{\gamma_D^+}{2}\pm\sqrt{e_D(\varepsilon +U)^2+T_C^2}
\label{eq:eigenenergies}
\eea
with $\gamma_\alpha^{\pm}\equiv (\Gamma_\alpha \pm\Gamma_\alpha')/2$ and  $e_\alpha(x)\equiv
(x+i\gamma_\alpha^-)/2$.
For $\Gamma'/\Gamma <
1+8T_C/\Gamma$ ($\varepsilon = U=$ 0) the energies (\ref{eq:eigenenergies}) have a nonvanishing real part, which provides the qubit oscillation 
frequency.
For $\Gamma'/\Gamma \ge
1+8T_C/\Gamma$ the eigenenergies are purely imaginary (see figure~\ref{fig1}), i.e., 
there is an abrupt  transition between an under- and an over-damped regime. 
The corresponding pure eigenstates are
\bea
\hat\rho_{Q\pm}=\frac{1}{2}
\cases{
\frac{1}{\gamma^-}\left(
\begin{array}{cc}
A_\pm & 4i\,T_C\\
-4i\,T_C & A_\mp
\end{array}
\right)
& for $\vert\gamma^-\vert> 4T_C$\\
\frac{1}{4T_C}\left(
\begin{array}{cc}
4T_C & iA_\pm \\
- iA_\mp & 4T_C
\end{array}
\right) & else
}
\eea
where $A_\pm\equiv \gamma^-\pm\sqrt{\vert(\gamma^-)^2-(4T_C)^ 2\vert}$,
$\gamma^\pm\equiv (\Gamma \pm \Gamma')/2$.
They are depicted in the Bloch sphere in figure~\ref{fig1}(c) for varying detection strength $\Gamma'/\Gamma$. 
For $\varepsilon=$ 0 and
$\vert\gamma_-\vert <4T_C$ the states "live" in the $x-y$ plane and for $\vert\gamma_-\vert >4T_C$ in the $y-z$ plane.
For a finite qubit bias $\varepsilon$  the Bloch vectors are not constrained to these planes anymore, as shown by the dashed curves in  figure~\ref{fig1}(c).


\begin{figure}[t]
\begin{center}
\includegraphics[width=.8\textwidth]{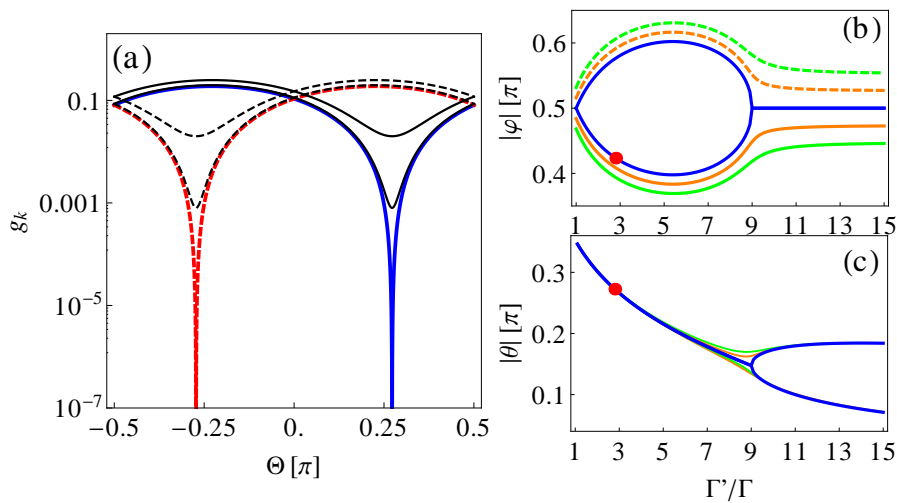}
\end{center}
\caption{(Color online) (a) Hilbert-Schmidt norm (\ref{eq:schmidt-hilbert}) vs. feedback parameter $\Theta$ 
for single qubit, $\Gamma'/\Gamma =$~2.8 and $\vert\varphi\vert
=$~0.4234$\pi$, $g_+$ (red, dashed) and  
$g_-$ (blue, solid). The shape of the curves around $g_k=$~0 provides the stabilization sensitivity on $\Theta$.
(b) Feedback angle $\varphi$ and (c) Feedback parameter $\Theta$ [see
(\ref{eq:singlefeed})] vs. coupling strength $\Gamma'/\Gamma$
required for 
perfect feedback stabilization $g_k=$~0. 
Different curves correspond to
$\varepsilon=$~0 (blue,
solid),
$\varepsilon=$~$-$0.1 (orange, solid), $\varepsilon=$~0.1 (orange,
dashed),
$\varepsilon=$~$-$0.2 (green, solid), $\varepsilon=$~0.2 (green, dashed).
Symbol \textcolor{red}{$\bullet$} indicates parameter set of red(dashed) and blue(solid) curves in (a).
For finite dissipation perfect stabilization fails, see thin black
curves in (a) $1/(\tau\Gamma) =$~0.02, 0.1.
Parameters: $U=$~0, $T_C/\Gamma=$~1.
}
\label{fig2}
\end{figure}


\noindent
{\em Feedback stabilization.} -- Now, we will study whether it is possible to stabilize these states by feedback.
As feedback operation on the qubit we introduce the following Hamiltonian
\bigskip
\bea
\hat{h}=\Theta\big[\sin{(\varphi)}\hat\sigma_x+\cos{(\varphi)}\hat\sigma_z\big],
\label{eq:singlefeed}
\eea
which allows for qubit rotations with feedback angle $\varphi$ and strength $\Theta$.

It turns out that for $\vert\gamma_-\vert >4T_C$ and $\varepsilon=$ 0 the corresponding eigenstates of the effective Hamiltonian
can be stabilized when $\varphi=\pi/2$.
The desired feedback strength $\Theta$ can be obtained analytically
and is provided in Ref.~\cite{KIE11}
for $\varphi=\pi/2$:
There are two distinct solutions for $\Theta$, where the state $\hat\rho_-$ 
provides the lower branch and   $\hat\rho_+$ provides the upper branch of figure~5 in \cite{KIE11}.

In order to stabilize the states $\hat\rho_\pm$ for $\vert\gamma^-\vert
<4T_C$ we need to adjust $\vert\varphi\vert\neq\pi/2$ (not considered in Ref.~\cite{KIE11}).
Then, the evaluation of the Hilbert-Schmidt norm
(\ref{eq:schmidt-hilbert}) reveals clear minima with 
$g_k =$~0 in that regime [figure~\ref{fig2}(a)].
The feedback angle $\varphi$ and strength $\Theta$ possess the same
absolute value for 
$g_\pm$ but different signs, their values for 
perfect stabilization are shown in figure~\ref{fig2}(b) and (c), respectively.

The single qubit can be purified in the
entire range of detection strength and for arbitrary qubit bias
$\varepsilon$.
However, additional dissipation that is not compensated by appropriate
feedback control actions, e.g., due to environmental charge fluctuations,
leads to the loss of stabilizability as indicated in figure~\ref{fig2}(a).

In the end of this section, we compare our studies with the work by Wang and Wiseman \cite{WAN01}, which 
deals with the purification of a two-level atom by optical feedback control.
It is based upon the unit-efficiency homodyne detection of the 
resonance fluorescence.
In contrast to our detection-based feedback scheme, where the control is triggered after a detection click is registered, their feedback Hamiltonian
is constantly applied to the system.
This leads to a different form of the unconditioned master equation (\ref{eq:master2}) and, consequently, to a distinct feedback behaviour.
A more detailed comparison of the homodyne-based with the detection-based feedback scheme in the context of entanglement generation in a quantum optical set-up 
can be found in Ref.~\cite{CAR08}.
However, in solid-state systems we are not aware of an electronic detection scheme yielding a formally equivalent description to 
homodyne-based feedback schemes in quantum optics \cite{WIS10}.
%


\subsection{\label{sec:entangle}Two interacting charge qubits: entanglement stabilization}

\begin{figure}[t]
\begin{center}
\includegraphics[width=.6\textwidth]{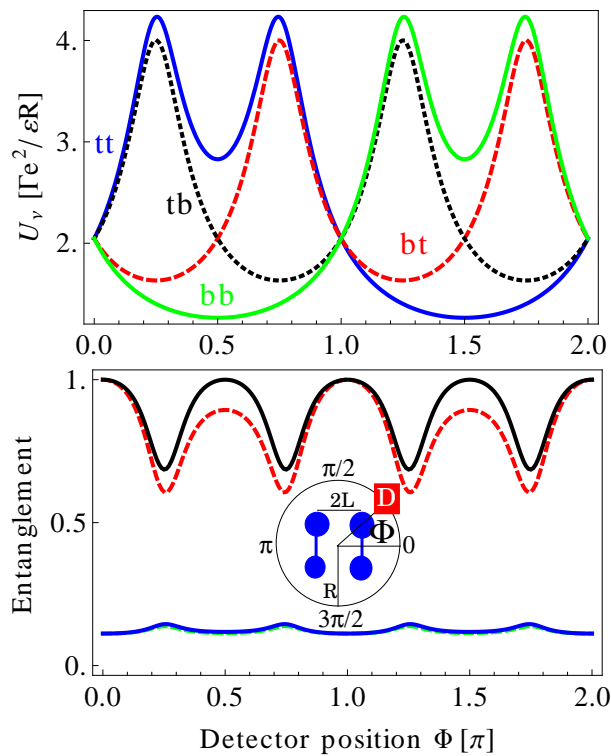}
\end{center}
\caption{(Color online) {\bf (Top)} Configuration interaction $U_\nu$ in dependence of the detector position $\Phi$ (see inset) 
for $L/R=$~1/2. In far-field ($L/R\to$~0) $U_\nu\to$~2 
independent of $\Phi$.
{\bf (Bottom)} Entanglement entropy of eigenstates of the effective
Hamiltonian. Other parameters are $U_0/\Gamma=$~0.5, 
$u/\Gamma=$~1, $T_1/\Gamma=T_2/\Gamma=$~1.
Complete bipartite entanglement (\ref{eq:entropy}) and detectability occurs at $\Phi_s=$~$(2n+1)\pi/2$ ($n\in\mathbf{Z}$) (black, solid curve) with Bell state $\ket{4}$ (\ref{eq:bell}).
{\bf (Inset)} Detector-qubit geometry.
}
\label{fig:detect}
\end{figure}

\noindent
{\it Model.} -- We consider an interacting bipartite system of two coupled qubits with the Hamiltonian
\bea
\label{eq:ham_coupled}
\hat{H}_Q&=&\hat{H}_1+\hat{H}_2+\hat{H}_{12}\\
\nn
\textrm{with}&&\quad \hat{H}_i=\frac{\varepsilon_i}{2}\hat\sigma_z^{(i)}+T_i\,\hat\sigma_x^{(i)},\quad
\hat{H}_{12}=\frac{u}{2}\,\hat\sigma_z^{(1)}\otimes\hat\sigma_z^{(2)},\nonumber
\eea
where $u\equiv u_\perp-u_\times$ [$u_\perp$ is the interaction between electrons in both top or bottom dots (see inset of figure~\ref{fig:detect}), 
$u_\times$ refers to the diagonal interaction], 
$\hat\sigma_z^{(i)}=\ket{i,t}\bra{i,t}-\ket{i,b}\bra{i,b}$ and $\hat\sigma_x^{(i)}=\ket{i,t}\bra{i,b}+\ket{i,b}\bra{i,t}$ ($i\in\{t,b\}$).
The corresponding eigenspectrum can be found in (\ref{eq:eigen_coupled}).

The qubit part of the Lindblad operators (\ref{eq:lindbladop}) reads here
\bea
\hat{B}_\alpha &=&\sqrt{\Gamma_\alpha^{(t,t)}}\ket{t,t}\bra{t,t}
+\sqrt{\Gamma_\alpha^{(t,b)}}\ket{t,b}\bra{t,b}+\sqrt{\Gamma_\alpha^{(b,t)}}\ket{b,t}\bra{b,t}\nn
&&+\sqrt{\Gamma_\alpha^{(b,b)}}\ket{b,b}\bra{b,b}.
\eea

\noindent
{\it Charge detector.} -- The system part of the detector coupling reads \mbox{$\hat{X}=\sum_j\frac{U_j}{2}\hat\sigma_z^{(j)}$}.
To be a bit more realistic in the following we will consider a specific geometry for the qubit system plus detector, which is shown in inset of figure~\ref{fig:detect}.
For simplicity we assume the four quantum dots of the qubit system placed on the corners of a square with edge length $2L$.
The SET detector is located on a circle surrounding the qubits with radius $R$; its position, then, is entirely determined by the angle $\Phi$ and $L/R$.
The derivation of the corresponding interaction strength $U_\nu$ can be found in \ref{sec:geometry}.

In order to provide a relation between  $U_\nu$ and $\Gamma^{(\nu )}$ we need to know the particular energy dependence of the SET tunnel rates $\Gamma$.
However, according to condition (\ref{eq:maxcond}) we assume that the energy dependence can be linearized:
\bea
\Gamma^{(\nu)}=\Gamma \frac{U_\nu}{U_0}
\eea
where $\nu\in\{\ket{t,t},\ket{t,b},\ket{b,t},\ket{b,b}\}$
and $\Gamma$ is the intrinsic energy-independent tunnel rate and $U_0$ provides the detector sensitivity.

In top figure~\ref{fig:detect} we show the particular dependence $U_\nu\propto\Gamma^{(\nu)}$ on the detector position $\Phi$.
At positions $\Phi=$~$n\pi$ ($n\in\mathbf{Z}$) there is no resolution on the qubit states since all $U_\nu$'s are equal; the detector is useless.
Of particular interest are the symmetric detector positions
$\Phi_s\equiv$~$(2n+1)\pi/2$, where it cannot discriminate between
state $\ket{\nu}=\ket{t,b}$ and $\ket{b,t}$.
This leads to a complete bipartite entanglement of one of the eigenstates of the effective Hamiltonian as shown in figure~\ref{fig:detect}; it is the Bell state $\ket{4}$ 
(\ref{eq:bell}).
Since we deal with the pure eigenstates of $\hat{H}_{\textrm{eff}}$
the standard definition 
by the von-Neumann entropy has been used:
\bea
S=-\textrm{Tr}[\rho_A\log_2 (\rho_A)]=-\textrm{Tr}[\rho_B\log_2 (\rho_B)]
\label{eq:entropy}
\eea
with $\rho_A=\textrm{Tr}_B[\rho]$, $\rho_B=\textrm{Tr}_A[\rho]$.
%

\begin{figure}[t]
\begin{center}
\includegraphics[width=.6\textwidth]{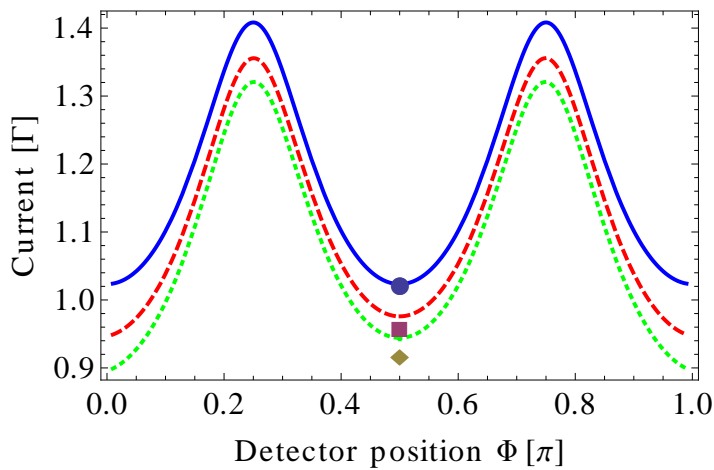}
\end{center}
\caption{
(Color online) Detector current in dependence of the detector position $\Phi$ without feedback.
A local minimum current is observed at  \mbox{$\Phi_s=$~$(2n+1)\pi/2$} ($n\in\mathbf{Z}$).
At these positions the SET detector current is bistable with one steady-state being the asymmetric Bell state $\ket{4}$ and the other represents 
a mixture of the remaining energy eigenstates. 
The respective currents are specified in (\ref{eq:bistab_currents}).
The curves correspond to different asymmetries $a=$~1 (blue, solid), 0.7 (red, dashed), 0.5 (green, dotted).
The symbols are: {\Large $\bullet$} ($\langle I\rangle_1$, $a$-independent), {\footnotesize $\blacksquare$} ($\langle I\rangle_2$ for $a=$~0.7),
{\footnotesize $\blacklozenge$} ($\langle I\rangle_2$ for $a=$~0.5).
Other parameters are  $U_0/\Gamma=$~0.5, 
$u/\Gamma=$~1, $T_1/\Gamma=T_2/\Gamma=$~1.
}
\label{fig:current}
\end{figure}

Before we will start with the discussions of entanglement stabilization by feedback it is worth to look at the behavior without feedback.
In \ref{sec:master} the quantum master equation in the energy
eigenbasis is written down completely.
Except of the symmetric detector positions $\Phi_s$ its steady state is unique and turns out to be a complete mixture -- the steady-state average current is
given by (The current is computed by the standard counting statistics method introduced in detail in, e.g., Ref.~\cite{SCH09c})
\bea
\langle I\rangle = e\frac{\Gamma^{(t,t)}+\Gamma^{(t,b)}+\Gamma^{(b,t)}+\Gamma^{(b,b)}}{16}.
\label{eq:current_coupledqb}
\eea
In contrast, the symmetric detector configuration yields a block structure of the Liouvillian superoperator (\ref{eq:super_coupled}) and, consequently, provides two steady states:
One is the Bell state $\ket{4}$ (\ref{eq:bell}) (The preparation
    of an entangled
    state by a current measurement alone has  been
    also reported by \cite{TRA06,WIL08}) and the other is the complete mixture of the remaining energy eigenstates.
The corresponding steady-state currents are ($T_1=T_2$)
\bea
\langle I\rangle_1 = e\frac{\Gamma^{(t,b)}}{4},\quad
\langle I\rangle_2 = e\frac{\Gamma^{(t,t)}+\Gamma^{(b,b)}+\Gamma^{(t,b)}}{12}.
\label{eq:bistab_currents}
\eea
For our specific choice of the $\Gamma^{(\nu )}$ in figure~\ref{fig:detect} we have  \mbox{$\Gamma^{(t,t)}+\Gamma^{(b,b)} = 2\Gamma^{(t,b)}$} 
so that all currents are equal: 
\mbox{$\langle I\rangle =\langle I\rangle_1=\langle I\rangle_2$} (blue, solid curve and symbol {\Large $\bullet$} in 
figure~\ref{fig:current}). 
Assuming some asymmetry $\Gamma^{(t,t)}+ a \Gamma^{(b,b)}\neq 2\Gamma^{(t,b)}$ for $a<$~1 due to, e.g., screening leads to 
\mbox{$\langle I\rangle\neq\langle I\rangle_1\neq\langle I\rangle_2$} [dashed ($a=$~0.7) and dotted  ($a=$~0.5) curves in 
figure~\ref{fig:current}]. The square/diamond symbol   corresponds to $\langle I\rangle_2$ for $a=$~0.7/$a=$~0.5, respectively.

In order to obtain one of these two states without feedback in the long-term limit one needs to initialize the system in the accompanying subspace \cite{SCH10}.
This is expected to be not only challenging but also vulnerable to any
slight perturbation that couples the two subspaces.
With the help of feedback it is possible to bypass this problem and
force the system into the 
desired state.


\begin{figure}[t]
\begin{center}
\includegraphics[width=.8\textwidth]{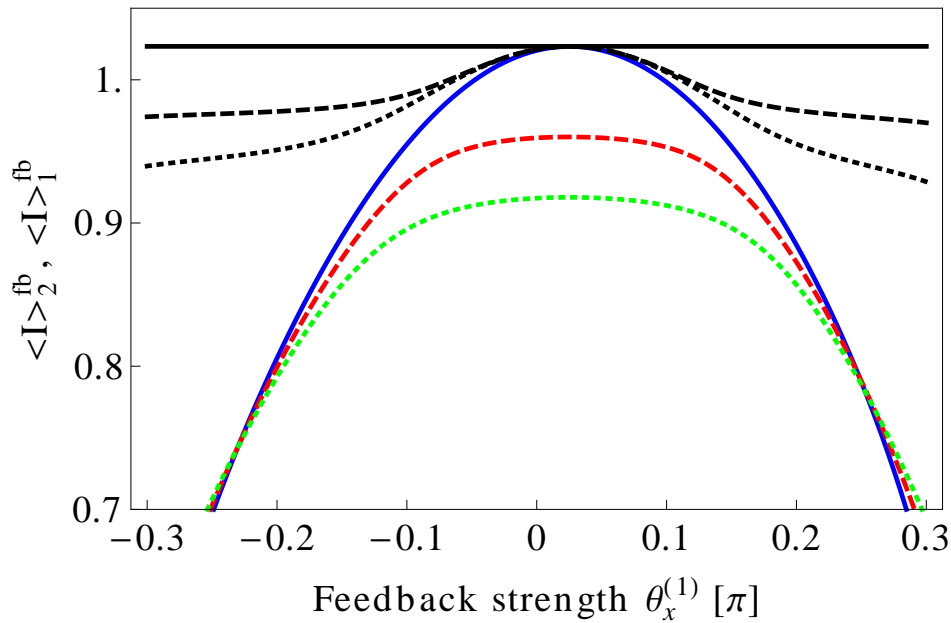}
\end{center}
\caption{
(Color online) Detector currents with feedback $\langle I\rangle_1^{\textrm{fb}}$ (black) and  $\langle I\rangle_2^{\textrm{fb}}$ (colored) in dependence of the feedback strength $\theta^{(1)}_x$ at detector positions \mbox{$\Phi_s=$~$(2n+1)\pi/2$} ($n\in\mathbf{Z}$).
For perfect stabilization the currents provide a maximum with respect to all feedback strengths $\theta_\gamma^{(i)}$ (here only shown $\theta^{(1)}_x$) 
with values given in (\ref{eq:bistab_currents}).
Therefore the SET current can be used to adjust the feedback strengths to obtain entanglement stabilization.
The curves correspond to different asymmetries $a=$~1 (solid), 0.7 (dashed), 0.5 (dotted).
Other parameters are  $L/R=$~0.5, $U_0/\Gamma=$~0.5, 
$u/\Gamma=$~1, $T_1/\Gamma=T_2/\Gamma=$~1.
}
\label{fig:currentFB}
\end{figure}


\noindent
{\it Feedback stabilization.} -- For the selection and stabilization of the maximally entangled state $\ket{4}$ we can apply the following local operation in 
(\ref{eq:feedback}), which acts on both qubits separately: 
\bea
\hat{U}
=e^{-i\hat{h}\delta t/\hbar}=e^{i \vec{n}_{1}\cdot\vec{\sigma}_1}\otimes e^{i \vec{n}_{2}\cdot\vec{\sigma}_2}
\label{eq:localuni}
\eea
where $\vec{n}_{i}=\sum_{\alpha=x,y,z}\theta^{(i)}_{\alpha}\mathbf{e}_\alpha$, $\vec{\sigma}_i=\sum_{\alpha=x,y,z}\hat\sigma^{(i)}_\alpha\mathbf{e}_\alpha$,
By  numerically minimizing the Hilbert-Schmidt norm $g_k$ (\ref{eq:schmidt-hilbert}) we find the following feedback parameters (in units of $\pi$)
\begin{center}
\begin{tabular}{|c|c|c|c|c|c|}
\hline
$\theta^{(1)}_x$ & $\theta^{(1)}_y$ & $\theta^{(1)}_z$ & $\theta^{(2)}_x$ & $\theta^{(2)}_y$ & $\theta^{(2)}_z$ \\
\hline
0.0252 & 0.1238 & 0.536 & $-$0.0206 & $-$0.1010 & $-$0.4373\\
\hline
\end{tabular}
\end{center}
which correspond to $g_4=$~0.
We remark that this set may not be unique because other minima with $g_4=$~0 might be found.
A stabilization is only achieved for the antisymmetric Bell state $\ket{4}$ at $\Phi_s$ and for $T=T_1=T_2$.
Remarkably, the parameters are independent of the values of the 
asymmetry $a$,
inter-qubit interaction $u$, qubit tunnel coupling $T$, and detector
sensitivity $U_0$.
As long as the inverse qubit-detector distance $L/R$ is larger than zero the feedback parameters are also independent of $L/R$.
Even though we succeeded with the stabilization of a maximally entangled state
we were not able to stabilize any other state at $\Phi_s$ of the
effective Hamiltonian and for $\Phi\neq\Phi_s$.
For that purpose we have used the most general form of an unitary transformation on SU(4) 
given by \cite{ZHA03}
\bea
\hat{U}=k_1\cdot k\cdot k_2,
\eea
consisting of a pulse sequence of two local operations $k_j$ (\ref{eq:localuni})
and a nonlocal operation
\bea
k=\prod_{\alpha=x,y,z} e^{i\theta_\alpha \hat\sigma_\alpha^{(1)}\otimes\,\hat\sigma_\alpha^{(2)}}.
\eea
%


The SET detector current can be used to monitor the effect of feedback, as shown in figure~\ref{fig:currentFB}.
For perfect entanglement stabilization with feedback strengths $\theta_\gamma^{(i)}$ given in the above table the bistable currents $\langle I\rangle_1$ and 
$\langle I\rangle_2$ provide maxima.
The corresponding current values are given in (\ref{eq:bistab_currents}).
Hence, by monitoring the SET current the feedback strengths can be adjusted very accurately in order to achieve perfect entanglement stabilization.

The idea of stabilizing entangled states by quantum-jump based feedback has been recently addressed by Carvalho and Hope \cite{CAR07,CAR08}.
In contrast to our all-electronic scheme, they study a pair of two-level atoms coupled to a single cavity mode; 
the feedback is triggered by a photodetector, which is not explicitly entering the calculations as in our studies.
Nevertheless there are some formal similarities between our works and findings, e.g., the occurence of the antisymmetric Bell state as steady state without feedback control, which are worth to
be further analyzed.


\section{\label{sec:conclude}Conclusions}

We have studied a method to stabilize pure states in interacting solid-state quantum systems based on 
electronic feedback triggered by single detector jumps.
This method facilitates the reverse engineering of quantum states.

In particular, a normal-conducting SET has been used as a realistic detector model in order to derive
a Lindblad quantum master equation for the coupled system of detector and quantum system.
This can be transferred into an effective Hamiltonian description, where the eigenstates of the 
effective Hamiltonian form the set of stabilizable states.
We have discussed the conditions under which the quantum-system state becomes pure even though the detector is in a 
transport state: (i) vanishing direct back-action from the detector and (ii) symmetric coupling in the single-electron transistor. 
This enabled us to formulate  the feedback  stabilization as an {\it inverse eigenvalue problem} -- the eigenstates and -energies belong to the effective 
Hamiltonian and the (feedback) superoperator will be determined.

We have illustrated the utility of our method in two examples:

Firstly, we applied it to the stabilization of pure states in a single charge qubit.
Beyond the studies in our Ref.~\cite{KIE11} here we were able to obtain the feedback operations, that purify the qubit for 
the whole range of detector-qubit coupling.
Thereby, the observed bifurcation is a property of the effective Hamiltonian spectrum and is expected to occur in more complex systems.
Additional dissipative sources destruct the effect of feedback stabilization. 

We have further 
studied the potential of the method to stabilize entanglement in a system of two interacting charge qubits.
Since the single-electron transistor couples capacitively this issue depends crucially on the geometry of the coupling between the detector and the qubit system.
We propose a realistic in-plane geometry, where at certain symmetric detector positions one of the eigenstates of the effective Hamiltonian is an asymmetric Bell state.
It turns out that this state can be stabilized within our feedback scheme, but all other (less-entangled) states are not stabilizable even with the help of 
general SU(4) operations.
We have demonstrated that by monitoring the detector current the feedback stabilization can be accurately tuned.

Some open questions may be addressed in future works:
Can one obtain some general statements on the existence and uniqueness of the solution of the inverse eigenvalue problem (\ref{eq:feedback})?
How can our general feedback scheme be transferred to set-ups with superconducting charge qubits or spin qubits used in recent experiments?


\ack

Financial support by the DFG (\mbox{BR 1528/8}, SFB 910, \mbox{SCH 1646/2-1}) and 
discussions with Howard Wiseman are gratefully acknowledged.

\appendix


\section{\label{sec:appendix1}Derivation of quantum master equation under feedback control}

Here, we will follow the derivation of an effective quantum master equation under the influence of single-jump feedback control
in Ref.~\cite{SCH12a}.
Along these lines we introduce measurement operators for the outcome jump-in (i), jump-out (o), no
jump (n) of electrons at the SET
\bea
\hat{M}_\textrm{i}(\Delta t)&=&\sum_\nu\sqrt{\Gamma^{(\nu )}_S\Delta t}\ket{\nu}\bra{\nu}\otimes
\hat{d}^\dagger =\hat{L}_S\sqrt{\Delta t},\nn
\hat{M}_\textrm{o}(\Delta t)&=&\sum_\nu\sqrt{\Gamma^{(\nu )}_D\Delta t}\ket{\nu}\bra{\nu}\otimes
\hat{d}=\hat{L}_D\sqrt{\Delta t},\nn
\hat{M}_\textrm{n}(\Delta t)&=&\sum_\nu\sqrt{1-\Gamma^{(\nu )}_S\Delta
  t}\ket{\nu}\bra{\nu}\otimes \hat{d}^\dagger \hat{d}\nn
&&+
\sum_\nu\sqrt{1-\Gamma^{(\nu )}_D\Delta t}\ket{\nu}\bra{\nu}\otimes(\mathbf{1}- \hat{d}^\dagger \hat{d}),\nn
\eea
which obeys the completeness relation 
$\hat{M}_\textrm{i}^\dagger (\Delta t)\hat{M}_\textrm{i}(\Delta t)
+\hat{M}_\textrm{o}^\dagger (\Delta t)\hat{M}_\textrm{o}(\Delta t)
+\hat{M}_\textrm{n}^\dagger (\Delta t)\hat{M}_\textrm{n}(\Delta t)=\mathbf{1}$.
For small $\Delta t$ we will do an expansion later on so that we need their action for $\Delta t =$~0:
\bea
\mathcal{M}_\textrm{i}(0)\hat\rho\quad\hat{=}&\quad& \hat{M}_\textrm{i}(0)\hat\rho \hat{M}_\textrm{i}^\dagger
(0)=\mathbf{0},\nn
\mathcal{M}_\textrm{o}(0)\hat\rho\quad\hat{=}&\quad& \hat{M}_\textrm{o}(0)\hat\rho \hat{M}_\textrm{o}^\dagger
(0)=\mathbf{0},\nn
\mathcal{M}_\textrm{n}(0)\hat\rho\quad\hat{=}&\quad& \hat{M}_\textrm{n}(0)\hat\rho \hat{M}_\textrm{n}^\dagger
(0)=\hat\rho,\nn
\mathcal{M}_\textrm{i}'(0)\hat\rho\quad\hat{=}&\quad& \frac{d}{d\Delta
  t}[\hat{M}_\textrm{i}(\Delta t)\hat\rho \hat{M}_\textrm{i}^\dagger
(\Delta t)]_{\Delta t=0}=\hat{L}_S\hat\rho\,\hat{L}_S^\dagger,\nn
\mathcal{M}_\textrm{o}'(0)\hat\rho\quad\hat{=}&\quad& \frac{d}{d\Delta
  t}[\hat{M}_\textrm{o}(\Delta t)\hat\rho \hat{M}_\textrm{o}^\dagger
(\Delta t)]_{\Delta t=0}=\hat{L}_D\hat\rho\,\hat{L}_D^\dagger,\nn
\mathcal{M}_\textrm{n}'(0)\hat\rho\quad\hat{=}&\quad& \frac{d}{d\Delta
  t}[\hat{M}_\textrm{n}(\Delta t)\hat\rho \hat{M}_\textrm{n}^\dagger
(\Delta t)]_{\Delta t=0}= - \frac{1}{2} \left\{\hat{L}_D^\dagger \hat{L}_D + \hat{L}_S^\dagger \hat{L}_S, \hat\rho\right\}.\nn
\eea
The feedback scheme is now defined by performing an instantaneous unitary transformation $\hat{U}_\alpha$ -- experimentally achieved by applying a $\delta$-pulse 
on the system Hamiltonian -- of the system density matrix whenever the detector generates a click. 
This leads us to the discrete iteration of the system density matrix
$\hat\rho (t+\Delta t)=\mathcal{P}(\Delta t)\hat\rho (t)$ with the effective propagator
\bea
\mathcal{P}(\Delta t)=e^{\mathcal{L}_0\Delta t}\mathcal{C}_S \mathcal{M}_\textrm{i}(\Delta
t)+
e^{\mathcal{L}_0\Delta t}\mathcal{C}_D \mathcal{M}_\textrm{o}(\Delta t)+e^{\mathcal{L}_0\Delta t}\mathcal{M}_\textrm{n}(\Delta t)
\eea
where $\mathcal{C}_\alpha\hat\rho\equiv e^{\mathcal{K}_\alpha}\hat\rho$
$\hat{=}$ $\hat{U}_\alpha\hat\rho\, \hat{U}_\alpha^\dagger$
and the Liouvillian super-operator $\mathcal{L}_0$ contains the system Hamiltonian and further un-monitored reservoirs.
We can use this propagator to eventually derive our effective master equation under unitary feedback control:
\bea
\frac{d}{dt}\hat\rho (t)&=&\lim_{\Delta t\to 0}\frac{\hat{\rho}(t+\Delta
  t)-\hat{\rho}(t)}{\Delta t}\nn
&=&\lim_{\Delta t\to 0}\frac{1}{\Delta
  t}[\mathcal{P}(\Delta t)-\mathbf{1}]\hat\rho (t)\nn
&=&\lim_{\Delta t\to 0}\frac{1}{\Delta
  t}\bigg\{\big[\mathcal{C}_S \mathcal{M}_\textrm{i}(0)+
\mathcal{C}_D \mathcal{M}_\textrm{o}(0)+\mathcal{M}_\textrm{n}(0)-\mathbf{1}\big]\nn
&&+
\Delta t\big[\mathcal{L}_0\mathcal{C}_S \mathcal{M}_\textrm{i}(0)+
\mathcal{L}_0\mathcal{C}_D
\mathcal{M}_\textrm{o}(0)+\mathcal{L}_0\mathcal{M}_\textrm{n}(0)\nn
&&+\mathcal{C}_S \mathcal{M}_\textrm{i}'(0)+
\mathcal{C}_D \mathcal{M}_\textrm{o}'(0)+\mathcal{M}_\textrm{n}'(0)\big]\bigg\}\hat\rho
(t)\nn
&=&\bigg[\mathcal{L}_0+\mathcal{C}_S \mathcal{M}_\textrm{i}'(0)+
\mathcal{C}_D \mathcal{M}_\textrm{o}'(0)+\mathcal{M}_\textrm{n}'(0)\bigg]\hat\rho
(t)\nn
&=&\bigg[\mathcal{L}_0+\mathcal{C}_S\mathcal{J}_S+
\mathcal{C}_D\mathcal{J}_D \bigg]\hat\rho
(t).
\eea
Correspondingly, jump super-operators in the no-feedback master
equation are supplemented by control operators to yield the feedback
master equation.


\section{\label{sec:appendix2}Coupled qubits without detector}

Let us assume unbiased qubits: $\varepsilon_i=$~0 $\forall i$.
Then the eigenspectrum of (\ref{eq:ham_coupled}) reads
\bea
e_{1/2}&=&\pm \Delta_+\,,\quad \ket{\psi_{1/2}}=\frac{a_+\pm a_-}{2}\ket{1}\pm\frac{a_+\mp a_-}{2}\ket{2},\nn
e_{3/4}&=&\pm \Delta_-\,,\quad \ket{\psi_{3/4}}=\mp\frac{b_+\pm b_-}{2}\ket{3}+\frac{b_+\mp b_-}{2}\ket{4},\nn
\label{eq:eigen_coupled}
\eea
with the Bell states
\bea
\ket{1}&=&\frac{1}{\sqrt{2}}\big[\ket{t,t}+\ket{b,b}\big],\quad
\ket{2}=\frac{1}{\sqrt{2}}\big[\ket{t,b}+\ket{b,t}\big],\nn
\ket{3}&=&\frac{1}{\sqrt{2}}\big[\ket{t,t}-\ket{b,b}\big],\quad
\ket{4}=\frac{1}{\sqrt{2}}\big[\ket{t,b}-\ket{b,t}\big],\nn
\label{eq:bell}
\eea
and $\Delta_\pm\equiv\sqrt{T_\pm^2+u^2/4}$, $T_\pm\equiv T_1\pm T_2$, $a_\pm\equiv\sqrt{1\pm T_+/\Delta_+}$, 
and $b_\pm\equiv\sqrt{1\pm T_-/\Delta_-}$.
For $u\to$~0 the eigenstates $\ket{\psi_i}$ are product states, whereas in the opposite limit $u\to\infty$ they become 
maximally entangled Bell states. 


\section{\label{sec:geometry}Coulomb interaction in the Detector-qubit geometry}

The configuration interaction in the geometry shown in figure~\ref{fig:detect} is simply given by
\bea
U_\nu =U_{1i,2j}=U_{1i}+U_{2j}=\frac{e^2}{\epsilon}\bigg(\frac{1}{l_{1i}}+\frac{1}{l_{2j}}\bigg),
\eea
with the elementary charge $e>1$, the dielectric constant $\epsilon$, and the distance between the detector and the qubit square corners
\bea
l_{ij}^2=2L^2+R^2-2\sqrt{2}LR\cos{(\theta_{ij})}
\eea
where
\bea
\theta_{1t}&=&\frac{3}{4}\pi-\Phi,\quad\theta_{1b}=\frac{5}{4}\pi-\Phi,\nn
\theta_{2t}&=&\Phi-\frac{\pi}{4},\quad\theta_{2b}=\Phi+\frac{\pi}{4}.
\eea
%


\section{\label{sec:master}Quantum master equation - Coupled qubits}

The quantum master equation for the coupled qubit system of section~\ref{sec:entangle}
in the energy eigenbasis (\ref{eq:eigen_coupled}) reads
\bea
\frac{d}{dt}\hat{\rho}(t)=\mathcal{L}\hat{\rho}(t)
\eea
with
\mbox{$\hat\rho
\equiv
\big(\boldsymbol{\rho}_{\textrm{p}},\boldsymbol{\rho}_{\textrm{ca}}
,\boldsymbol{\rho}_{\textrm{cb}}\big)^T$}, where
\mbox{$\boldsymbol{\rho}_{\textrm{p}}\equiv\big(\boldsymbol{\rho}_{11},\boldsymbol{\rho}_{22},
\boldsymbol{\rho}_{33},\boldsymbol{\rho}_{44}\big)$},
\mbox{$\boldsymbol{\rho}_{\textrm{ca}}\equiv\big(\boldsymbol{\rho}_{12},\boldsymbol{\rho}_{23},
\boldsymbol{\rho}_{13},\boldsymbol{\rho}_{24},\boldsymbol{\rho}_{34},\boldsymbol{\rho}_{14}\big)$},
\mbox{$\boldsymbol{\rho}_{\textrm{cb}}\equiv\big(\boldsymbol{\rho}_{21},\boldsymbol{\rho}_{32},
\boldsymbol{\rho}_{31},\boldsymbol{\rho}_{42},\boldsymbol{\rho}_{43},
\boldsymbol{\rho}_{41}\big)$},
\mbox{$\boldsymbol{\rho}_{ij}=\big(\rho_{ij}^{(0)},
\rho_{ij}^{(1)}\big)$},
and the Liouvillian super-operator
\bea
\mathcal{L}\equiv\frac{1}{2}
\left(
\begin{array}{ccc}
\label{eq:super_coupled}
\mathcal{L}_{\textrm{pop}} & \mathcal{L}_{\textrm{pc}} &
\mathcal{L}_{\textrm{pc}}^* \\
\mathcal{L}_{\textrm{pc}}^\dagger & \mathcal{L}_{\textrm{cc}} & \mathbf{0}\\
\mathcal{L}_{\textrm{pc}}^{\textrm{T}} & \mathbf{0} &
\mathcal{L}_{\textrm{cc}}^*
\end{array}
\right).
\eea
Its sub-matrices for the (pop)ulation sector, the coupling sector between population and coherences (pc), and the coherences sector (cc) are given by  
\bea
&&\mathcal{L}_{\textrm{pop}}\equiv
\left(
\begin{array}{cccc}
A_{12}^a & B_a &  D_{12} & \color{red}{C_{12}} \\
B_a & A_{21}^a &  C_{21} & \color{red}{D_{21}} \\
D_{12} & C_{21} & B_{12}^b &\color{red}{B_b}\\
\color{red}{C_{12}} &\color{red}{D_{21}} & \color{red}{B_b} &
\mathbf{\color{blue}{A_{21}^b}} 
\end{array}
\right),\nonumber
\eea
\bea
\mathcal{L}_{\textrm{pc}}\equiv\left(
\begin{array}{cccccc}
E_{21}^a & I_{21}^a & K_{12,12}^a &   \color{red}{L_{12}^a} & 
\color{red}{J_{21,21}} & \color{red}{H_{12,12}^a} \\
(E_{12}^a)^* &  - H_{21,21}^a &  -L_{21}^a  & \color{red}{K_{21,21}^a}
& 
\color{red}{-J_{12,12}} & \color{red}{I_{12}^a} \\
- J_{21,12} & H_{21,12}^b &  K_{12,12}^b & \color{red}{L_{21}^b} & 
\color{red}{(E_{12}^b)^*} & \color{red}{-I_{21}^b} \\
\color{red}{J_{12,21}} & \color{red}{-I_{12}^b} &
\color{red}{-L_{12}^b}  & \color{red}{K_{21,21}^b}
& \color{red}{E_{21}^b} & \color{red}{H_{12,21}^b}\\
\end{array}
\right),\nonumber
\eea
\bea
\mathcal{L}_{\textrm{cc}}\equiv\left(
\begin{array}{cccccc}
F_a &   -L_{21}^a & -H_{21,12}^a  & \color{red}{I_{12}^a}
&\color{red}{M} & \color{red}{K_{21,12}^a}\\
-L_{21}^a & N_{12}^* & P_{21}^a  & \color{red}{P_{21}^b}&
\color{red}{- L_{21}^b} &   \color{red}{O} \\
- H_{21,12}^a &  P_{21}^a  & S_{12} & \color{red}{O}&\color{red}{-
  I_{21}^b} & \color{red}{P_{12}^b} \\
\color{red}{I_{12}^a} & \color{red}{P_{21}^b} & 
\color{red}{O} & S_{21}^* &  - H_{21,21}^b & P_{12}^a \\
\color{red}{M} &  \color{red}{- L_{21}^b} & 
\color{red}{- I_{21}^b} & - H_{21,21}^b & F_b & K_{12,21}^b \\
\color{red}{K_{21,12}^a} &  \color{red}{O} 
& \color{red}{P_{12}^b} & P_{12}^a &  K_{12,21}^b & N_{21} 
\end{array}
\right).\nonumber
\eea
The corresponding 2$\times$2 sub-matrices are defined as
\bea
A_{ij}^x&=&\big(-x_1^2\gamma_i^+-x_2^2\gamma_j^+\big)\mathbf{1}+\big(x_1^2\hat\gamma_i^++x_2^2\hat\gamma_j^+\big)^2\hat{\sigma}_x',\nn
B_x&=&[x_1x_2(\hat\gamma_1^+-\hat\gamma_2^+)]^2\hat{\sigma}_x',\nn
C_{ij}&=& (a_1b_2\hat\gamma_i^-+a_2b_1\hat\gamma_j^-)^2\hat{\sigma}_x',\nn
D_{ij}&=&(a_1b_1\hat\gamma_i^--a_2b_2\hat\gamma_j^-)^2\hat{\sigma}_x',\nn
I_{ij}^x&=&x_1x_2(a_2b_2\hat\gamma_i^--a_1b_1\hat\gamma_j^-)(\hat\gamma_1^+-\hat\gamma_2^+)\,\hat{\sigma}_x',\nn
L_{ij}^x&=&x_1x_2(a_1b_2\hat\gamma_i^-+a_2b_1\hat\gamma_j^-)(\hat\gamma_1^+-\hat\gamma_2^+)\,\hat{\sigma}_x',\nn
J_{ij,kl}&=&(a_1b_2\hat\gamma_i^-+a_2b_1\hat\gamma_j^-)(-a_1b_1\hat\gamma_k^-+a_2b_2\hat\gamma_l^-)\,\hat{\sigma}_x',\nn
M&=&(a_1b_1\hat\gamma_1^--a_2b_2\hat\gamma_2^-)(a_1b_1\hat\gamma_2^--a_2b_2\hat\gamma_1^-)\,\hat{\sigma}_x',\nn
O &=&  - a_1 a_2 b_1 b_2 (\hat\gamma_1^+ - \hat\gamma_2^+)^2\,\hat{\sigma}_x',\nn
%
E_{ij}^x&=&-\bigg\{\pm\frac{1}{2}x_1x_2(\gamma_1^+-\gamma_2^+)+4i\big[T_\pm (x_1^2-x_2^2) - u\,x_1x_2)\big]\bigg\}\mathbf{1}\nn
&&\pm x_1x_2(x_i^2\hat\gamma_1^++x_j^2\hat\gamma_2^+)(\hat\gamma_1^+-\hat\gamma_2^+)
\,\hat{\sigma}_x',\nn
H_{ij,kl}^x&=&-\frac{1}{2}(a_1b_2\gamma_i^-+a_2b_1\gamma_j^-)\mathbf{1}+(a_1b_2\hat\gamma_i^-+a_2b_1\hat\gamma_j^-)(x_1^2\hat\gamma_k^++x_2^2\hat\gamma_l^+)\,\hat{\sigma}_x',\nn
K_{ij,kl}^x&=&\frac{1}{2}(a_1b_1\gamma_i^--a_2b_2\gamma_j^-)\mathbf{1}+(a_2b_2\hat\gamma_j^--a_1b_1\hat\gamma_i^-)(x_1^2\hat\gamma_k^++x_2^2\hat\gamma_l^+)\,\hat{\sigma}_x',\nn
F_x&=&-\frac{1}{2}\big[32ix_1x_2T_\pm + 8iu(x_1^2-x_2^2)+\Gamma/2 \big]\mathbf{1}\nn
&&+(x_1^2\hat\gamma_1^++x_2^2\hat\gamma_2^+)(x_1^2\hat\gamma_2^++x_2^2\hat\gamma_1^+)\,\hat{\sigma}_x',\nn
P_{ij}^x&=&\frac{1}{2}\bigg\{8 i[ (x_1^2 - x_2^2) T_+ -  x_1 x_2\,u]\mp x_1 x_2 (\gamma_1^+ -\gamma_2^+)\bigg\}\mathbf{1} \nn
&& \pm x_1 x_2 ( \bar{x}_i^2 \hat\gamma_2^+ + \bar{x}_j^2 \hat\gamma_1^+ 
  ) (\hat\gamma_1^+ - \hat\gamma_2^+) \,\hat{\sigma}_x',\nonumber
\eea
\bea
N_{ij}&=&\bigg\{-8 i \big[a_1 a_2 T_++ b_1 b_2 T_-+(a_1^2 
      b_1^2 - a_2^2 
      b_2^2) u\big]\nn
&& - 2( a_1^2 b_2^2 \gamma_j^+ + a_2^2b_1^2 \gamma_i^+)-
(a_1^2b_1^2 + a_2^2 b_2^2)\Gamma\bigg\}\mathbf{1}\nn
&&+\bigg\{(a_i^2 \hat\gamma_2^+ + 
   a_j^2 \hat\gamma_1^+) (b_j^2 \hat\gamma_2^+
   + b_i^2\hat\gamma_1^+)\bigg\}\,\hat{\sigma}_x',\nn
S_{ij}&=&\bigg\{-8 i [a_1 a_2 T_+ - b_1b_2T_- - (a_2^2 b_1^2 - a_1^2b_2^2)u] \nn
&&-2(a_2^2b_2^2 \gamma_i^+ + a_1^2 b_1^2\gamma_j^+) - (a_2^2 b_1^2 + a_1^2 b_2^2)\Gamma\bigg\}\mathbf{1}\nn
&& + \bigg\{(a_i^2 \hat\gamma_2^+ + 
   a_j^2 \hat\gamma_1^+) (b_i^2 \hat\gamma_2^+ + 
   b_j^2 \hat\gamma_1^+)\bigg\}\,\hat{\sigma}_x',\nonumber
\eea
with $\hat{\sigma}_x'\equiv e^{i\chi}\ket{t}\bra{b}+\ket{b}\bra{t}$, $\Gamma\equiv\Gamma^{(tt)}+\Gamma^{(bb)}+\Gamma^{(tb)}+\Gamma^{(bt)}$, $\gamma_1^\pm\equiv\Gamma^{(tt)}\pm\Gamma^{(bb)}$, 
$\gamma_2^\pm\equiv\Gamma^{(tb)}\pm\Gamma^{(bt)}$, $\hat\gamma_1^\pm\equiv\sqrt{\Gamma^{(tt)}}\pm\sqrt{\Gamma^{(bb)}}$, 
$\hat\gamma_2^\pm\equiv\sqrt{\Gamma^{(tb)}}\pm\sqrt{\Gamma^{(bt)}}$, $a_1\equiv\frac{a_++ a_-}{2\sqrt{2}}$, $a_2\equiv\frac{a_+- a_-}{2\sqrt{2}}$,
$b_1\equiv\frac{b_++ b_-}{2\sqrt{2}}$, $b_2\equiv\frac{b_+- b_-}{2\sqrt{2}}$.

It is observed, that 
at $\Phi=(2n+1)\pi/2$ the Liouvillian superoperator (\ref{eq:super_coupled}) decomposes into block structure since $\gamma_2^-=\hat\gamma_2^-=$~0 and $b_2=$~0 ($T_-=$~0);
vanishing sub-matrices are marked by red.
The $A_{21}^b$-block becomes decoupled and belongs to the $\ket{4}$ Bell
state (\ref{eq:bell}) for $u\gg$~1, which therefore represents the steady state of this block.
The steady-state of the remainder is a complete mixture.
The aim of the feedback control is the unique selection and stabilization of the maximally entangled state $\ket{4}$.
%


\section*{References}

\bibliographystyle{/usr/share/texmf/bibtex/bst/base/unsrt}
\bibliography{/home/gerold/references}

\end{document}